# Severe Language Effect in University Rankings
## Particularly Germany and France are wronged in citation-based rankings

Anthony F.J. van Raan, Thed N. van Leeuwen, and Martijn S. Visser
*Centre for Science and Technology Studies (CWTS)*
*Leiden University*

*Abstract*

*We applied a set of standard bibliometric indicators to monitor the scientific state-of-arte of 500 universities worldwide and constructed a ranking on the basis of these indicators (Leiden Ranking 2010). We find a dramatic and hitherto largely underestimated language effect in the bibliometric, citation-based measurement of research performance when comparing the ranking based on all Web of Science (WoS) covered publications and on only English WoS covered publications, particularly for Germany and France.*

The WoS data system covers a number of journals in non-English languages, particularly in German and in French. Publications in these non-English language journals are counted as part of the output of countries, but they generally have a very low impact because only a few scientists outside Germany, Austria and Switzerland are able to read German, and a similar situation for French. Thus, these non-English publications will 'dilute' considerably the average impact of countries such as Germany, Austria and France. This is particularly the case for the more application-oriented fields such as clinical medicine and engineering, and also for the social sciences and humanities. Because clinical medicine represents a considerable part of the entire scientific output of a country, this language effect directly influences the overall impact position of the university (van Leeuwen, Moed, Tijssen, Visser and Van Raan 2001).

This effect has serious consequences for all citation-per-paper based rankings, but also for citations per staff rankings. The most striking result is a too low position of most German and France universities in these rankings, particularly universities with a medical school. As citations play an important role in university rankings such as the Times Higher Education, QS, Shanghai (ARWU) and the Leiden Ranking, all these rankings suffer from this language effect.

Thus, for ranking and comparison with countries such as the US, UK and all countries where almost all WoS-covered publications are written in English, calculation of bibliometric indicators based on only-English publications is the only fair procedure.

We show results in Figure 1 for in total 69 German and French universities. We see that the 'English-only' citation impact is systematically higher than the 'all publications' impact, and the ranking position based on 'English-only' citation impact is systematically lower (thus, 'better') than based 'all publications' impact, with a few exceptions due to differences in inter-group changes of ranking positions. In Figure 2 the language effect is shown in a slightly different way for the 25 German universities with the highest citation impact. A comparison of the rankings for the 500 largest universities in the world based on all WoS publications as well as on English-only publications is given in our ranking website (Leiden Ranking 2010).

One can expect that similar effects will occur when using Scopus data and that this effect will seriously aggravate by the extension of the coverage of the WoS and Scopus with national language journals.

## References

van Leeuwen, T.N., H.F. Moed, R.J.W. Tijssen, M.S. Visser, and A.F.J. van Raan (2001). Language biases in the coverage of the Science Citation Index and its consequences for international comparisons of national research performance. *Scientometrics* 51(1), 335-346.

Leiden Ranking 2010: see http://www.socialsciences.leiden.edu/cwts/products-services/leiden-ranking-2010-cwts.html

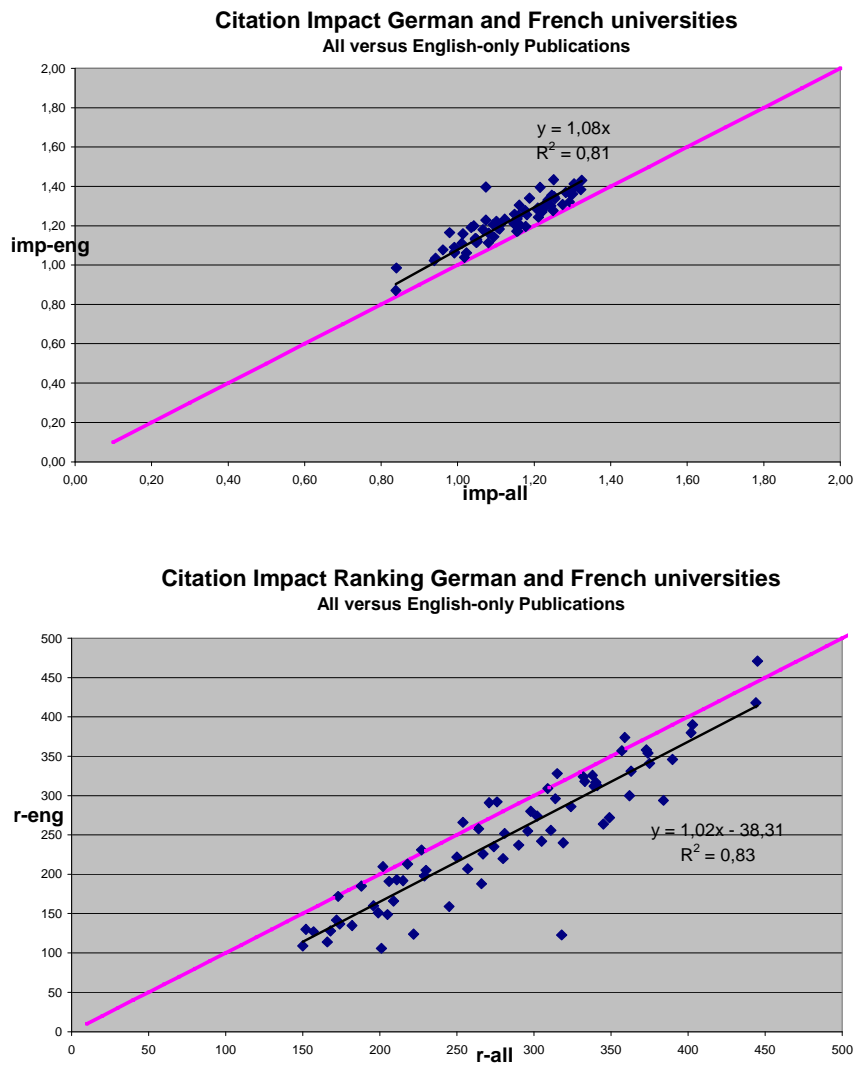

Fig 1. Language effect in citation impact of all German and French universities. Upper figure shows the difference in impact (numerical values of the field-normalized citation score) in the case of all WoS and only-English WoS publications. Lower figure shows the difference in citation impact ranking position in the case of all WoS and only-English WoS publications.

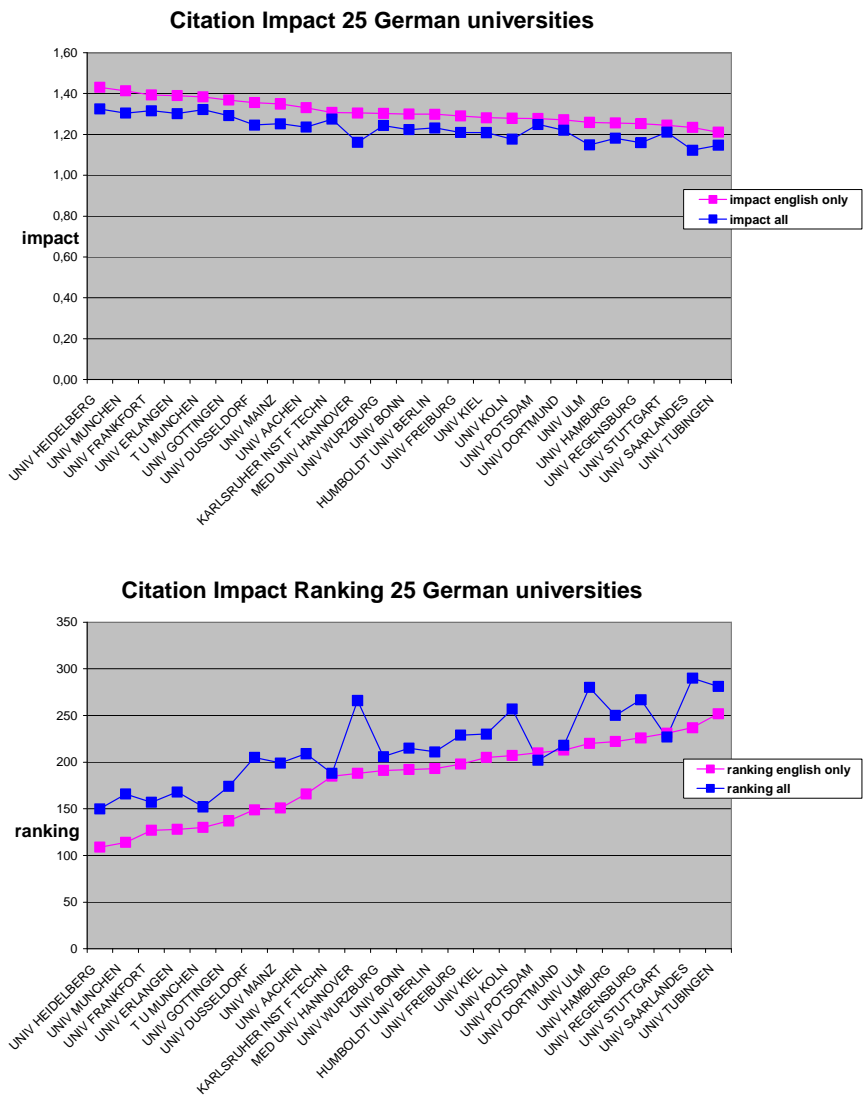

Fig 1. Language effect in citation impact of the 25 highest impact German universities. Upper figure shows the difference in impact (numerical values of the field-normalized citation score) in the case of all WoS and only-English WoS publications. Lower figure shows the difference in citation impact ranking positions in the case of all WoS and only-English WoS publications.